\documentclass[%
aip,
amsmath,amssymb,
reprint,%
]{revtex4-1}

\usepackage{graphicx}
\usepackage{amsmath}
\usepackage{amsfonts}
\usepackage{latexsym}
\usepackage{amsbsy}
\usepackage{epstopdf}
\usepackage{bm}%

\usepackage{color}

\usepackage{ulem} 

\newcommand{\dyadic}[1]{{#1}
\setbox0=\hbox{$\scriptstyle\leftrightarrow$}
   \setbox2=\hbox{$#1$}
   \dimen0=.5\wd0 \advance\dimen0 by-.5\wd2
   \advance\dimen0 by-\wd0
   \kern\dimen0
{^{\hbox{$\scriptstyle\leftrightarrow$}}}}

\begin{document}

\title{
Electromagnetically induced transparency based Rydberg-atom sensor for quantum voltage measurements}
\thanks{Publication of the U.S. government, not subject to U.S. copyright.}
\author{Christopher L. Holloway}
\affiliation{National Institute of Standards and Technology, Boulder,~CO~80305, USA}
\email{christopher.holloway@nist.gov}
\author{Nikunjkumar Prajapati}
\affiliation{National Institute of Standards and Technology, Boulder,~CO~80305, USA}
\affiliation{Department of Physics, University of Colorado, Boulder,~CO~80305, USA}
\author{John Kitching}
\author{Jeffery A. Sherman}
\author{Carson Teale}
\author{Alain R\"ufenacht}
\author{Alexandra B. Artusio-Glimpse}
\author{Matthew T. Simons}
\affiliation{National Institute of Standards and Technology, Boulder,~CO~80305, USA}
\author{Amy K. Robinson}
\affiliation{Department of Electrical Engineering, University of Colorado, Boulder,~CO~80305, USA}
\author{Eric B. Norrgard }
\affiliation{National Institute of Standards and Technology, Gaithersburg, Maryland 20899, USA}
\date{\today}

\begin{abstract}
We investigate the Stark shift in Rydberg rubidium atoms through electromagnetically induced transparency for the measurement of direct current (dc) and 60~Hz alternating current (ac) voltages. This technique has direct applications to atom-based measurements of dc and ac voltage and the calibration of voltage instrumentation. We present experimental results for different atomic states that allow for dc and ac voltage measurements ranging from 0~V to 12~V.
A Rydberg atom-based voltage standard could become an alternative calibration method with more favorable size, weight, power consumption, and cost compared to the more precise Josephson voltage standard.
In this study, we also demonstrate how the voltage measurements can be utilized to determine the atomic polarizability for the Rydberg states. The Rydberg atom-based voltage measurement technology would become a complimentary method for dissemination of the voltage scale directly to the end user.

\end{abstract}

\maketitle

\section{Introduction}

Rydberg atoms (atoms with one or more electrons excited to a very high principal quantum number $n$ \cite{gal}) in conjunction with electromagnetically-induced transparency (EIT) techniques have been used to successfully detect and fully characterize radio frequency (RF) electric (E) fields \cite{r14, gor1, sed1, holl1, holl2}. In these applications, the E-fields are detected using EIT both on resonance as Autler-Townes (AT) splitting and off-resonance as ac Stark shifts.  This approach has the capability of measuring amplitude \cite{gor1, sed1, holl1, holl2, tan1, r5, gor2, gor3}, polarization \cite{sed2, access}, and phase \cite{sim3, jing1} of the RF field and various applications are beginning to emerge. These include E-field probes \cite{holl1, holl2, gor2} traceable to the International System of units (SI), power-sensors \cite{holl5}, spectrum analyzers \cite{army}, angle-of-arrival detection\cite{aoa}, and receivers for communication signals (AM/FM modulated and digital phase modulation signals) \cite{song1, meyer1, holl6, cox1, holl4, anderson2, deb3, holl7}.

\begin{figure}[!t]
 \centering
\scalebox{.18}{\includegraphics*{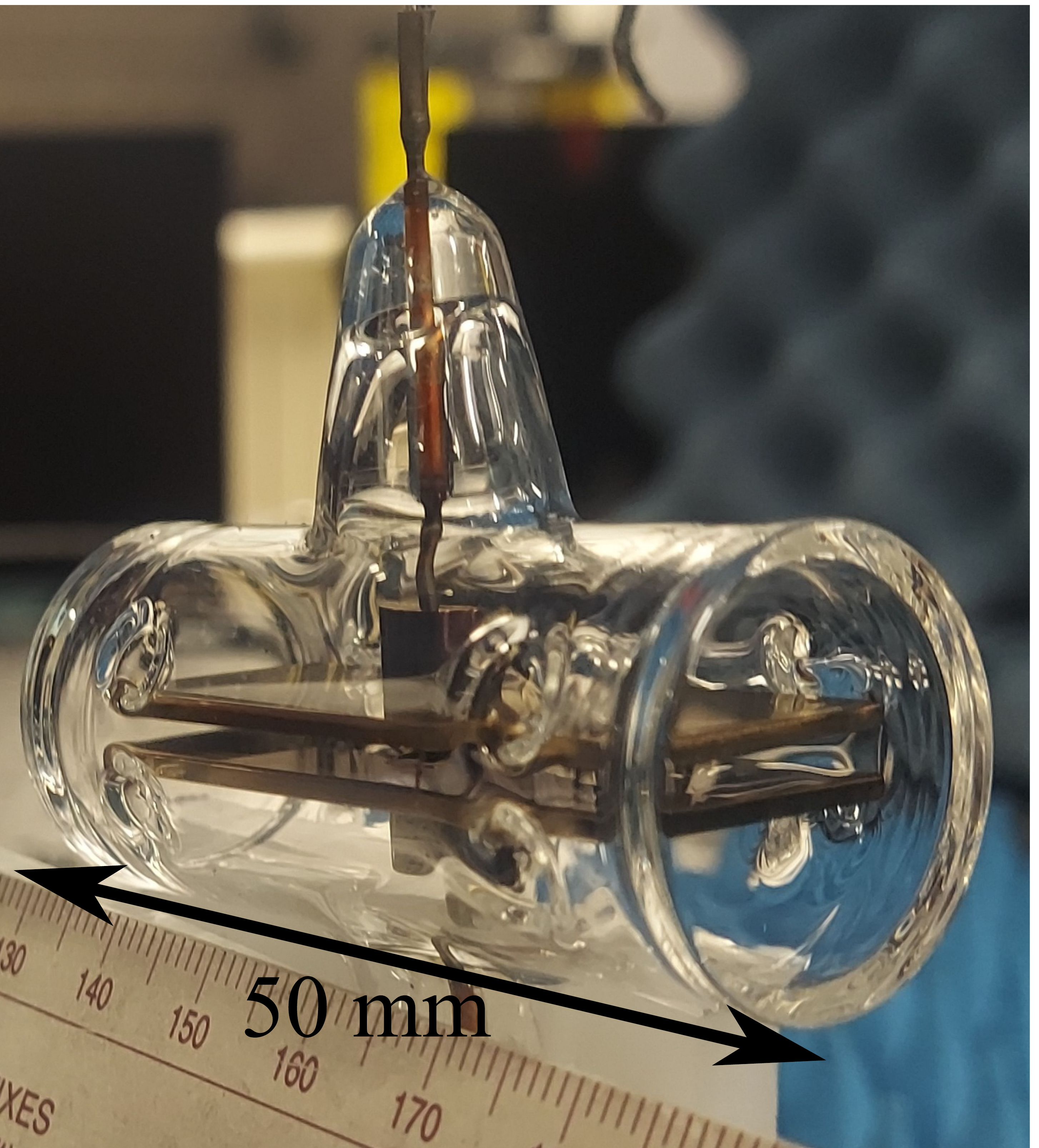}}
\hspace{5mm}
\caption{Cylindrical vapor cell with stainless-steel parallel plates. The vapor cell is 50~mm in length and has a outside diameter of 25~mm. The plates are rectangular in shape with width of 18~mm, length of 45~mm, and placed $\approx2$~mm apart.}
\label{cell}
\end{figure}

In this manuscript, we investigate the use of Rydberg-atom sensors to develop a quantum-based voltage standard. Here, we measure the voltage induced between two parallel plates embedded in an atomic vapor cell (see Fig.~\ref{cell}) by measuring the Stark shifts in the atomic spectra of Rydberg atoms. By collecting a series of measurements of the Stark shift for different voltages, we can make a calibration curve and demonstrate a voltage standard.  We discuss various issues that must be considered in order for Rydberg-atom-based-sensors to accurately and reliably function as voltage measurement standards.

The  primary voltage standard implemented for the realization of the SI volt is based on the Josephson effect. Arrays of Josephson junctions, when cooled to cryogenic temperatures and biased on their quantum locking range transduce an accurately synthesized microwave frequency source into fixed or digitally programmable voltages \cite{jose}. Comparisons between such voltage standards demonstrate accuracies at the level of 1 part in $10^{10}$ with about 100~s of averaging (see, e.g. \cite{refa}). Scaling up the voltage output of Josephson-junction based standards requires either increasing the number of junctions in the array, or increasing the microwave driving frequency, or both \cite{refb}. With some exceptions (see, e.g. \cite{refc}), 10~V is the commonly adopted dc reference for the primary realization of the voltage scale. Commercial instruments (like voltage calibrators) used for the routine voltage calibration process and voltage dividers implemented to expand the range to high voltages typically have an accuracy on the order of 1 part in $10^6$~ \cite{refd}. Due to their large cost ($\gtrsim ~$\$$300$k) Josephson voltage standards limit their deployment to national metrological institutes (NMIs) or primary calibration laboratories. 

Zener diodes are the most commonly used voltage references that are implemented with commercial instrumentation, ranging from high performance secondary voltage standard ($>~$\$10k) \cite{z1,z2} to the 0.1~\% accuracy handheld multimeters ($\approx~$\$100). All Zener-diode voltage references drift with time and require periodic calibration, that involves a comparison with a more accurate voltage reference or another verification of the manufacturer specifications. Zener-diode references are sensitive to temperature, humidity, mechanical stress and shock, and power supply noise \cite{z1}.


Rydberg atom sensors, that rely on the atom's predictable response to electric field to measure voltage could become an alternative --- in terms of size, weight, power and cost (SWaP-C) --- to Zener voltage references and, with lesser accuracy, to Josephson voltage standards. For example, 
polarizability measurements with low-lying excited states of alkaline-earth atoms show total inaccuracy at the level of 2 parts in $10^{5}$ ~\cite{refe, reff}. Due to a large dipole moment, Rydberg atomic states have high sensitivity to electric fields, high intrinsic response bandwidth, and can be produced by laser excitation in large quantities within simple vapor cells at or near room temperature. In principle, the atomic response can be made free from environmental coupling, which would make Rydberg atom sensors intrinsically stable, eliminating the need of subsequent periodic external calibrations. 


In this paper, we discuss the various nuances of Rydberg-atom-based voltage measurements [e.g., the effects of energy level crossing and non-parallel electrodes (causing inhomogeneous fields)]. While similar measurements have been performed before \cite{das1, das2, adams2, oster}, our study focuses on the aspects that will need to be addressed and controlled for a Rydberg-atom sensor to be used as a voltage measurement standard. We also demonstrate two methods for measuring ac voltage amplitude at 60~Hz. AC Stark effect measurements have been performed in Ref.~\cite{das1}, where it was shown that indium tin oxide electrodes where limited to 40~kHz and nickel electrodes operated down to dc.
Our cells achieve dc and low frequency sensing with stainless steel plates. Finally, we discuss how accurate measurements of the Stark shifts and the plate separation can lead to accurate estimates of the polarizability for Rydberg states.




\section{Voltage Measurements}

A Rydberg atom-based dc (or 60\,Hz) voltage measurement is performed by observing the dc (quasi-dc) differential Stark shifts in Rydberg states, which is given by \cite{friedrich}
\begin{equation}
\Delta=\frac{\alpha(f)}{2}\,\,E^2 \,\,\, ,
\label{stark}
\end{equation}
where $\Delta$ is the Stark shift (in units of Hz), $E$ is the applied electric field (in units of V/m) with frequency $f$ (in units of Hz), and $\alpha(f)$ is the  polarizability [in units of Hz/(V$^2$/m$^2$)].  For Rydberg state $i$, the polarizabiltiy may be expressed as
\begin{equation}\label{eq:alpha}
    \alpha_i(f) =\sum_j \frac{\vert \mu_{ij}^z\vert^2}{h} \Big(\frac{1}{f_{ij}-f-\frac{i}{2}\Gamma_{j}}+\frac{1}{f_{ij}+f+\frac{i}{2}\Gamma_{j}}\Big).
\end{equation}
Here, the summation includes all states $j$ which are coupled to $i$ by electric field, $\mu_{ij}^{z}$ is the $z$ component of the dipole matrix element, $f_{ij}$ is the transition frequency, and $\Gamma_{j}$ is the spontaneous decay rate (in units of Hz) of state $j$, respectively. Table~\ref{t1} gives the calculated dc polarizability $\alpha(0)$ for the atomic states used in the experiments presented in this paper using Eq.~\eqref{eq:alpha} and dipole matrix elements $\mu_{ij}$ taken from the Alkali Rydberg Calculator Python package \cite{ARC2020}.  This package calculates dipole matrix elements  using the Coulomb approximation (CA) and the quantum defects given in Ref.\,\cite{han2006}.   Reference \cite{yerokhin2016} compared the accuracy of  polarizablities calculated using the CA  to calculations with the more detailed Dirac-Fock with core potential approach.  For the 28S to 47S Rydberg states considered here, the theoretical polarizability values calculated using the CA are accurate to better than 1\,\%.  The accuracy of the theoretical values exceed the experimental accuracy for these states, which is typically on the level of a few percent \cite{osullivan1985,osullivan1986}. Because $f_{ij}\,\gg\,60\,$Hz, $\alpha(f=60\,\rm{Hz})\approx\alpha_f(0)$. To the precision given in Table~\ref{t1} ($< 0.1$\,\%), $\alpha$ is identical for $f=0$ and $f=60$\,Hz. 

If the E-field is generated by applying a voltage across two parallel plates with separation $d$ (neglecting fringing field, see discussion in section \ref{sec:fringing}), then the E-field is $E=V/d$. By substituting this into equation Eq.~(\ref{stark}), the voltage can be found by measurement of the Stark shift:
\begin{equation}
V=d\,\,\sqrt{\frac{2\,\Delta}{\alpha}} \,\,\, .
\label{starkV}
\end{equation}
From this expression, we see that there are three sources of uncertainties in this measurement. There are two measured quantities in this expression (the plate separation $d$ and the Stark shift $\Delta$) and one calculated quantity (the atomic polarizability, $\alpha$). $\alpha$ can also be determined experimentally.  

The measurement of $\Delta$ can be related to \textit{in situ} laser spectroscopy of the hyperfine structure, contributing a relative precision of  $10^{-7}$ or better~\cite{refhh}.
Relative uncertainty in $d$ of $10^{-4}$ is possible with gauge-block construction \cite{refi} or at $< 2 \times 10^{-5}$ with interferometric techniques \cite{refe, reff}. Typically, poor knowledge of $\mu_{ij}$ limits the calculation of $\alpha$ to the $10^{-3}$ level, though in some cases $\alpha$ is experimentally determinable by other means \cite{refj}. Even lacking absolute knowledge, the quantity $d / \sqrt{\alpha}$ can be calibrated to the extent that $d$ remains stable. For a given device, Equation 3 is cast as:
\begin{equation}
V = C_\text{cal} \sqrt{\Delta}
\label{starkV2}
\end{equation}
where $C_\text{cal} = 2d / \sqrt{\alpha}$ is determined by comparison to a higher accuracy voltage standard, or by use of several Rydberg states with various $\alpha$ of lower uncertainty.


The dc and 60 Hz measurements will share a common $C_\text{cal}$. Other sources of uncertainty arise from fringing fields near the perimeter of the electrode plates,  non-idealities in plate geometry and orientation, and spectroscopic features due to interfering transitions.




\begin{table}
\caption{Theoretical calculation of dc and 60~Hz ac polarizabilities for $^{85}$Rb. Note that the dc and 60~Hz ac polarizabilities are essentially the same for the digits shown}
\label{t1}
\begin{tabular}{|| c | c  ||}
 \hline
 \multicolumn{2}{|c|}{Calculated Polarizability [Hz/(V/m)$^2$] } \\
 \hline  \hline
 Rydberg state & $\alpha$: {\bf dc}  and  {\bf 60~Hz ac}  \\
 \hline \hline
$28S_{1/2}$ & 84.2   \\
 \hline
$40S_{1/2}$ & 1058.1 \\
 \hline
$47S_{1/2}$ & 3275.7    \\
 \hline
\end{tabular}
\end{table}

\section{Experimental Setup}

In these experiments, we generate EIT in rubidium ($^{85}$Rb) atomic vapor and measure the frequency shift in the EIT signal as a function of an applied voltage across two parallel plates.  The experimental apparatus is shown in Fig.~\ref{cell}. It consists of a cylindrical-vapor cell of length 50~mm and diameter of 25~mm. Inside the cell are two stainless-steel parallel plates with a nominal separation of $d = 2$~mm. 
We had two of the same type of cells manufactured and we are able to determine that the plate separation for the two cells were approximately 2.19~mm (referred to as ``Cell~1'') and 2.11~mm (referred to as ``Cell~2''), which is within the uncertainty for the plate separation stated by the manufacturer. While the data shown in this paper are for Cell~1, we do comment on the data from Cell~2. The plates in both cells are not perfectly parallel to one another, which 
causes field inhomogeneities across the plates. We discuss the effect this has on the EIT spectra and voltage measurement in Section~\ref{sec:fringing}.

\begin{figure}
    \centering
 \scalebox{.26}{\includegraphics{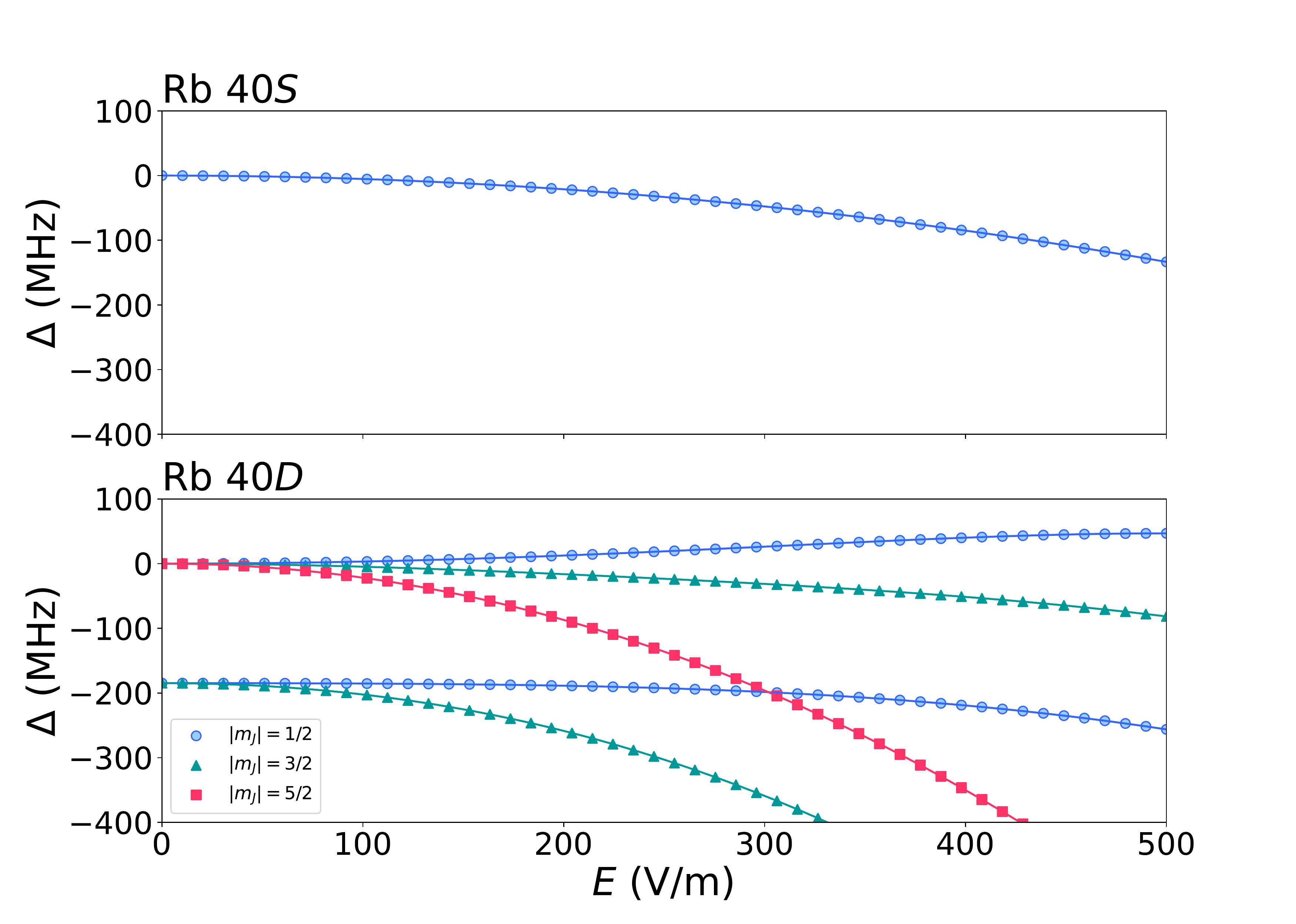}}
    \caption{Calculated dc Stark shift $\Delta_f$ in $^{85}$Rb as a function of electric field $E$ for $40S$ and $40D$.}
    \label{map}
\end{figure}

Using a Rydberg-atom EIT measurement scheme for measuring Stark shifts involves exciting the atoms to some Rydberg state.  States with zero orbital angular momentum ($S$ states) exhibit a less complicated spectrum as a function of applied electric field. Fig.~\ref{map} shows the calcualted Stark shift of the $^{85}$Rb $40S_{1/2}$ state (Fig.\,\ref{map}a) and $40D_{m_j}$ state (Fig. \,\ref{map}b).  Compared to $S$ states, the Stark spectrum of $D$ states is typically complicated by the fine structure, resulting in many more possible energy level crossings (ELC) than seen in the $S$ states. We therefore focus on $S$ states as they are typically well described by Eq.\,\ref{stark}. However, as we will see below, for large applied voltages, ELCs do appear in the $S$-state spectra and can influence the measurements.

\begin{figure}[!t]
\centering
\scalebox{.27}{\includegraphics*{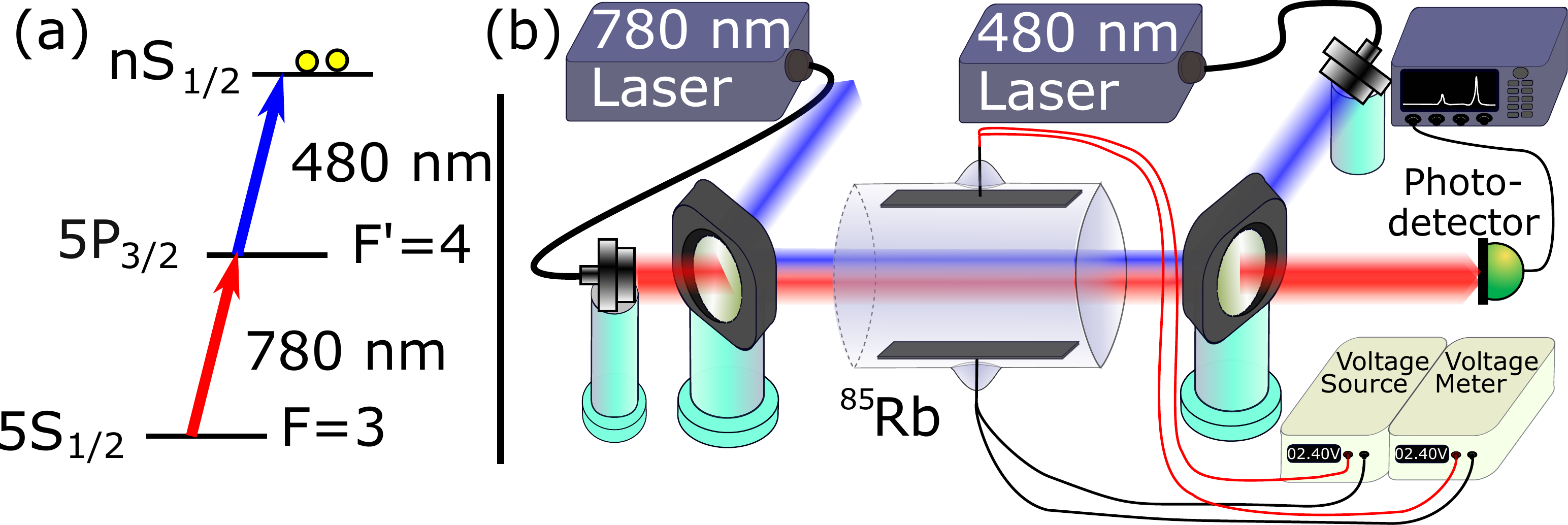}}\\
\caption{(a) Level diagram depicting EIT coupling the 5S$_{1/2}$,F=3 state to a nS$_{1/2}$ Rydberg state through the 5P$_{3/2}$,F'=4 intermediate state. (b) Experimental setup for the voltage measurement and the three-level EIT scheme.}
\label{setup}
\end{figure}

The experimental setup and the atomic levels used are depicted in Fig.~\ref{setup}, which consists of a 780~nm probe laser, a 480~nm coupling laser, a photodetector connected to an oscilloscope, a voltage source, a voltage meter, and the vapor cell shown in Fig.~\ref{cell} filled with $^{85}$Rb atomic vapor. We use a three-level EIT scheme to generate Rydberg atoms [see Fig.~\ref{setup}(a)] which corresponds to the $^{85}$Rb $5S_{1/2}$ as the ground state, $5P_{3/2}$ as the intermediate state, and a Rydberg state of n$S_{1/2}$ state. In these experiments we use $n=47$, $n=40$, and $n=28$.
The probe laser is locked to the D2 transition (\mbox{$5S_{1/2}(F=3)$~--~$5P_{3/2}(F=4)$} or wavelength of \mbox{$\lambda_p=780.24$~nm} \cite{stackrb}).  To produce an EIT signal, we apply a counter-propagating coupling laser with $\lambda_c \approx 480$~nm and scan it across the $5P_{3/2}$-$nS_{1/2}$ Rydberg transition. We use a lock-in amplifier to enhance the EIT signal-to-noise ratio by modulating the coupling laser amplitude with a 37~kHz square wave. This removes the background and isolates the EIT signal. 

A voltage source is connected to the electrodes on the outside of the vapor cell shown in Fig.~\ref{cell}.  These electrodes penetrate the cell and are connected to the two parallel plates. The EIT signal frequency shift (the Stark shift) is measured for different applied voltages. The voltage meter is also connected to the electrodes in order to monitor the applied voltage. Both the voltage sources and voltage meter were calibrated before the experiments and the absolute voltage accuracy of any reading is 0.5~mV. In these experiments, the optical beams and the electric fields  between the two plates are co-linearly polarized. The probe laser was focused to a full-width at half maximum (FWHM) of 80~$\mu$m with a power of 3.6~$\mu$W, and the coupling laser was focused to a FWHM of 110~$\mu$m with a power of 70~mW.



\section{Experimental Data}

\begin{figure}[t!]
	\centering
	\scalebox{.45}{\includegraphics*{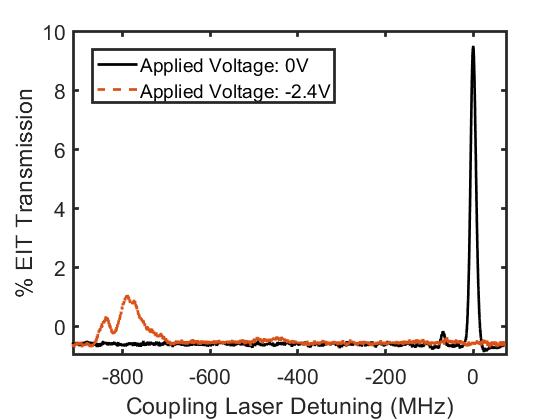}}\\
	\caption{Plot of \% probe EIT transmission relative to the probe absorption plotted against the coupling laser detuning. The black-solid curve is the transmission trace for the 40S Rydberg state when no voltage is applied across the plates. The red-dashed curve is the transmission trace for the 40S Rydberg state when a potential of 2.4 V is placed on the plates. 
	}
	\label{fig:sample}
\end{figure}

The EIT signals for two different applied voltages are shown in Fig.~\ref{fig:sample}. These results are for the 40$S_{1/2}$ state as the coupling laser is scanned. To increase the signal to noise ratio, we gather 20 oscilloscope traces and average them after accounting for laser drift using an external reference EIT signal.  The two different traces correspond to the zero voltage case (black-solid trace) and the case with 2.4 V applied (red-dashed trace). For no applied voltage, we see two peaks, the main EIT peak at \mbox{$\Delta_c/2\pi=0$~MHz} and another at \mbox{$\Delta_c/2\pi=-75.6$~MHz}, which corresponds to the hyper-fine structure transition $5P_{3/2}(F=3)\rightarrow 40S_{1/2}$. Since the coupling laser is scanned, the separation between these two peaks is 75.63~MHz. The detuning of the observed peak is the adjusted hyper-fine splitting determined by accounting for the doppler mismatch between the probe and coupling lasers with \mbox{$120.96(\frac{\lambda_p}{\lambda_c}-1)$~MHz}. Here, $\lambda_p$ and $\lambda_c$ are the wavelengths of the probe and coupling lasers, and 120.96~MHz is the hyper-fine structure separation between 5P$_{3/2}$(F=3) and 5P$_{3/2}$(F=4) \cite{stackrb}. The separation between the main EIT peak and the peak corresponding to the hyper-fine structure is used to calibrate the laser detuning and the frequency for the measured Stark shift ($\Delta$).


When a voltage is applied, the EIT peak shifts according to Eq.~(\ref{starkV}).  When 2.4~V is applied, we see the EIT peak shift to around -800~MHz (the red trace). We see that the width of the EIT signal is broader than when no voltage is applied. The broadening is due to inhomogeneous fields that occur across the two laser propagation paths. This is discussed in more detail later. We also see asymmetries and two peaks in the observed EIT signal. These two peaks and asymmetries are a result of both ELCs and inhomogeneous fields.



\begin{figure*}
	\centering
	\scalebox{.65}{\includegraphics*{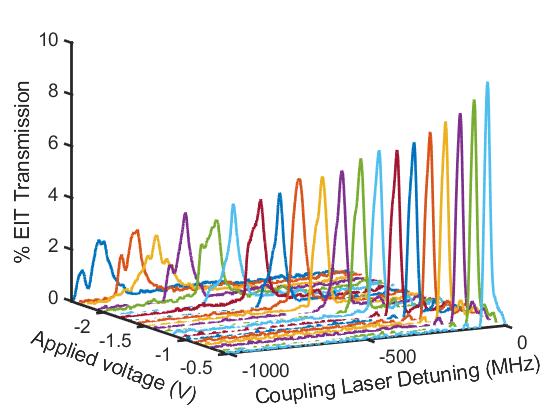}}\\
	\caption{EIT signals (for scanning the coupling laser) showing Stark shifts for various applied voltages for 40S$_{1/2}$.}
	\label{eitlines}
\end{figure*}

 Fig.~\ref{eitlines} shows several EIT signals for several applied voltages. This series of EIT signals allow us to observe several interesting features.  Besides Stark shifts and line broadening, we see other features that appear in the EIT lines, including additional structure in the EIT signal that start to appear. The structure becomes complicated for higher applied voltages. These structures stem from two sources. First, these structures are due to the ELC beginning to appear in the spectra. 
Secondly, the source of these structures (that eventually develops into double peaks for higher voltages) are a result of inhomogeneities in the field across the laser beam propagation path.  The inhomogeneities in the field are mainly caused by (1) non-parallel plates, (2) imperfections in the surface of the electrodes due to attaching lead to the plates [as our current plates have a dimple (and discontinuities) in the very center where the leads are mounted], and (3) fringing fields at the edge of the plates.
To confirm the plates are not parallel, we used a microscope to measure the plate separation at the four corners. 
We measured the electrode separation in each cell at each of the four corners ten times. For the cell, these measurements (and standard uncertainties) were: 2.306(11)~mm, 2.105(7)~mm, 2.278(7)~mm, and 2.103(14)~mm. From this data, we see that the plates are tilted (uniformly along the length of the cell) with respect to another. 

We see the EIT signals broaden with applied voltage as seen in Fig.~\ref{eitlines}.  This broadening is likely due to a non-uniform field across the plates along the laser propagation path.  Furthermore, additional broadening may be present from local charge distributions. These are likely from ionization of the Rydberg atoms through collision~\cite{das1}.
Additionally, these E-field in-homogeneities are amplified by the fact that the Stark effect is quadratic for non-degenerate energy levels~\cite{oster}. Imperfect plate geometries have other ramifications as we discuss below. 
The second cell (Cell~2 mentioned above) had not only tilts in the plate, but had a twist in the plate. While not shown here, the measurements from the second cell showed even more broadening and the appearance of double peaks in the EIT lines.

\begin{figure}
	\centering
	\scalebox{.50}{\includegraphics*{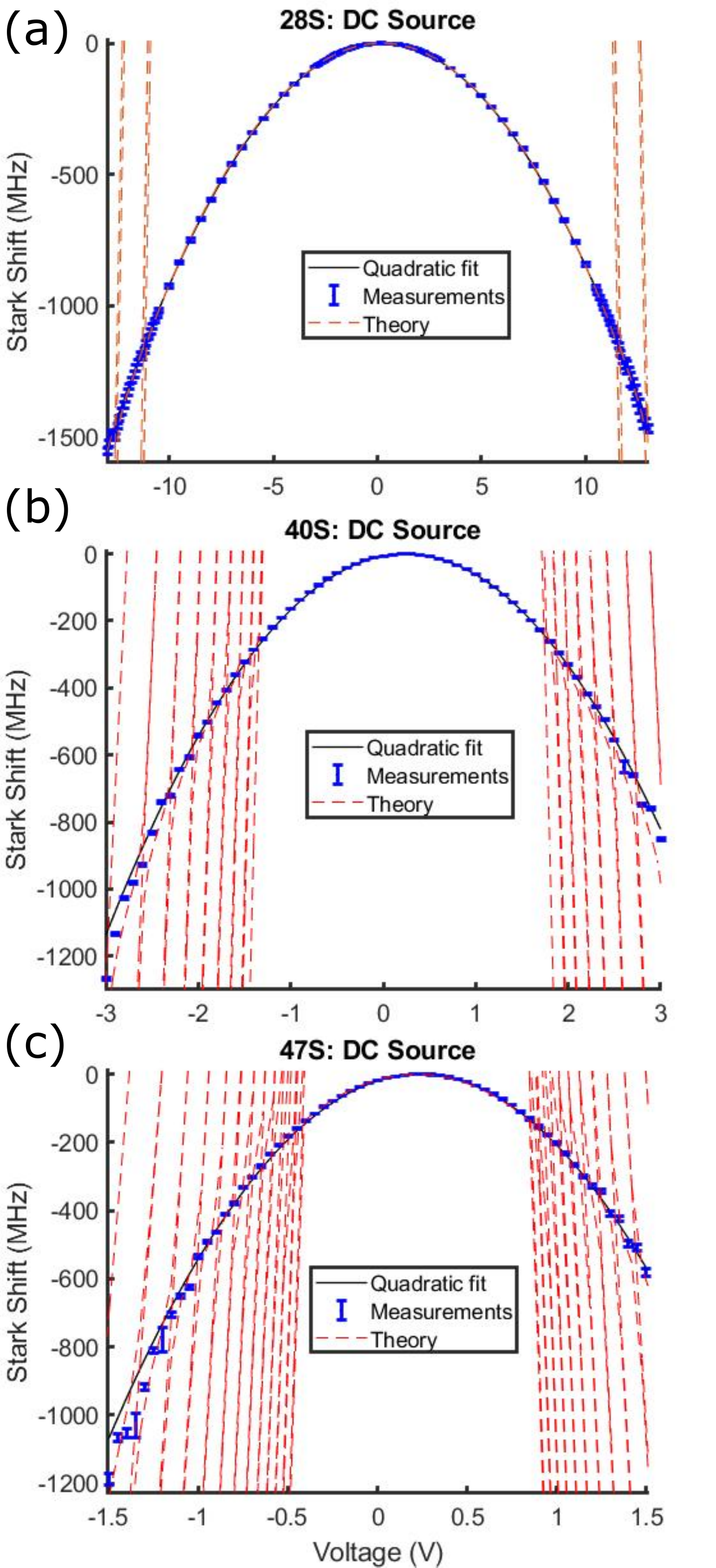}}\\
	\caption{(a)-(c) correspond to the measured Stark shift for 28S, 40S, and 47S Rydberg states with a dc source (blue errorbars). Also shown are the curve fits using the calibration factor found by eq.~(\ref{fit1}). The dashed-red lines are from theory and when these lines become vertical indicate the position of the ELCs.}
	\label{fig:data_shift}
\end{figure}

We extract the measured Stark shifts (corresponding to the location of the maximum of the shifted EIT peak) and plot them against the applied voltage. This was done for the three Rydberg states mentioned previously (28S, 40S, and 47S), shown by Fig.~\ref{fig:data_shift}. The error bars in these plots correspond to the standard deviation from six sets of data. In this data, we can also identify that the Stark shift is not zero for zero applied voltage. This offset is possibly due to excess charge in both vapor cells.
Grounding the two plates did not remove this excess charge. The voltage offset comes from various sources, including (1) impurities in the metal causing a build up of charge which cannot be removed and (2) ionization caused by the 480~nm coupling laser. It is also possible that this is a result of a galvanic potential. Whatever the case may be this offset can be accounted for.

In order to determine the offset, we fit to the equation
\begin{equation}
\Delta =\frac{1}{C^2_{cal}}\left(V-V_o\right)^2 \,\,\, .
\label{fit1}
\end{equation}
where $V_o$ is the offset voltage caused by the excess charge. We found the voltage offset is 235$~\pm$~6~mV for this cell. These fits were within a $95~\%$ confidence interval. Through the fit, we are also able to find the calibration factor for each cell as measured using the three Rydberg states, given in Table.~\ref{tab:cals}.  Fig.~\ref{fig:data_shift} shows these fits for the three states. In determining the calibration factor $C_{cal}$ care must be taken to ensure the ELCs do not influence its value. The effect of the ELCs are discussed in the next two sections.

\begin{table}
\caption{Experimentally obtained calibration constants. The uncertainty is the standard deviation of the calibration constant obtained from the 6 sets of data.}
\label{tab:cals}
\begin{tabular}{|| c | c  ||}
 \hline
 \multicolumn{2}{|c|}{Calibration Factors} \\
 \hline  \hline
 Rydberg state & C$_{Cal}$ (V/$\sqrt{Hz}$) \\
 \hline \hline
 $28S_{1/2}$ Cell 1 &  $(339 \pm 2.7)\cdot10^{-6}$  \\
 \hline
 $40S_{1/2}$ Cell 1 &  $(96.7 \pm 0.2)\cdot10^{-6}$    \\
 \hline
 $47S_{1/2}$ Cell 1 &  $(53.8 \pm 0.3)\cdot10^{-6}$   \\
 \hline
\end{tabular}
\end{table}

In addition to this, $C_{cal}$ can be used to either estimate the polarizability $\alpha (0)$ or to determine the plate separation. In the first case, the calibration factor found through the fit of the Rydberg atom Stark shift measurements and the independently measured plate separations can be used to quantify the polarizability $\alpha_e$. In the second case, the calibration factor and theoretically calculated polarizability can be used to find the plate separation. 
In the first case, if the plate separation can be independently determined, then $C_{cal}$ can be used to experimentally determine $\alpha_e$ by the following [obtained by comparing eq.~(\ref{starkV}) to eq.~(\ref{starkV2})]:
\begin{equation}
    \alpha_e=2d^2/C_{cal}^2 \,\,\, .
    \label{calalpha}
\end{equation}

Using this expression and the average plate separations for the four corners (measurement with a microscope) given above ($d=2.194$~mm for the cell), the estimated $\alpha_e(0)$ for each state and each cell is given in Table~\ref{talpha}. In this table we also show the percent error (\mbox{$\Delta\%$=100*[$\alpha_e(0)-\alpha(0)]/\alpha(0)$]}, where $\alpha(0)$ is given in Table~\ref{t1}). 
Better control of the plate separation would allow for more accurate determination of $\alpha_e(0)$.

\begin{table}
\caption{Experimentally obtained $\alpha_e(0)$, percentage error from calculated $\alpha(0)$, and experimentally obtained plate separations. The uncertainty is the standard deviation from 6 sets of data..}
\label{talpha}
\begin{tabular}{|| c | c | c | c ||}
 \hline
 Rydberg state & $\alpha_e(0)$ & $\Delta\%$ & Plate separation (mm) \\
 \hline \hline
 $28S_{1/2}$ Cell 1 &  $ 83.95 \pm 1.4$    & 0.29 &   $2.20  \pm 0.019$  \\
 \hline
 $40S_{1/2}$ Cell 1 &  $ 1032.43 \pm 4.4$   & 2.4 &  $2.21 \pm 0.005$ \\
 \hline
  $47S_{1/2}$ Cell 1 &  $ 3328.59 \pm 33.2$   & 1.6 & $2.16  \pm 0.013$  \\
 \hline

\end{tabular}
\end{table}

\section{Energy Level Crossings}

\begin{figure}
	\centering
\scalebox{.45}{\includegraphics{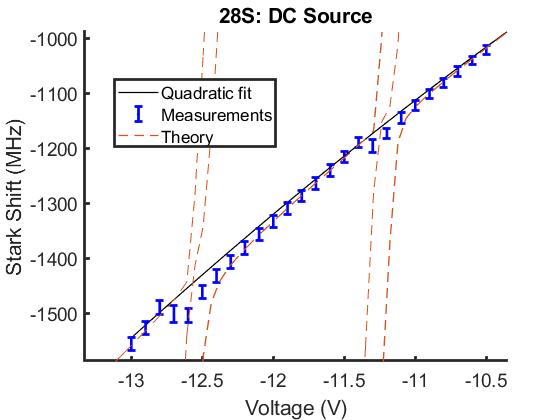}}
	\\
	\caption{Expanded view of the ELC's for the 28$S_{1/2}$ state. We can observe the measurement (blue error bars) deviations from the quadratic nature (solid black line) of the dc Stark effect at these locations. Also shown is the theory (red dashed line) that accounts for the ELC's and matched the data much better.}
	\label{zoomcrossover}
\end{figure}

The measured data in Fig.~\ref{fig:data_shift} were corrected for the voltage offset and compared to theoretical calculations for the Stark shift. The theoretical Stark maps were obtain by using the Alkali Rydberg Calculator Python package \cite{ARC2020}. These comparisons are shown in Fig.~\ref{fig:data_shift}, where the theoretical model accounts for ELCs from nearby Rydberg states. We show plots for the Stark shift as a function of applied $V$. In the theoretical curves, the $E$ field was determined using $E=V d$ (where we use $d=2.19$~mm). We can see good consensus between the experimental data and theoretical model. In particular, on closer inspection, the locations of the ELCs can be seen in the measured data. This is further illustrated in Fig.\ref{zoomcrossover}, where we show an expanded view of the ELC for the 28$S_{1/2}$ state spectra, shown in Fig.~\ref{fig:data_shift} (a). In this figure, the level crossings are easily seen as a splitting in the spectra around -11.2~V and -12.5~V.  On this note, the Stark shifts corresponding to the ELCs can be used to find the calibration factor as well.



After obtaining a consensus on the calibration constant through multiple measurements, we can utilize the cells for measuring dc voltages. However, the ELCs can limit the maximum voltage that can be accurately measured since they cause deviations from a perfect quadratic dependence of the frequency on the applied voltage. For a given state, it would be best to avoid voltage levels where the first ELCs appear.  For example, this occurs at approximately 12~V, 1.5~V, and 0.6~V, for $28S_{1/2}$, $40S_{1/2}$, and $47S_{1/2}$ respectively. The quadratic behavior depicted in eqs.~(\ref{stark}) and (\ref{starkV}) fails near the ELCs.  As such, using the eq.~(\ref{starkV2}) and measured Stark shifts will need to be limited to voltages well below the first ELC. So for large voltage measurements Rydberg states with low $n$ should be used, while for small voltages Rydberg states with high $n$ are more suitable (due to the higher sensitivity to weak fields for high $n$). 
Alternatively, choosing the electrode plate separation, or manufacturing a variety of plate separations in a cell (with care and attention to fringing fields and geometrical accuracy) allows a given voltage to produce a field appropriate for accurate measurement by a particular Rydberg state $n$.

\section{Subtleties to Calibration}
\label{sec:fringing}

In practice, a calibration must account for or correct three non-idealities: non-uniformity (spatial regions where the electric field between electrodes departs from the relationship E=V/d, from fringing fields or imperfections in plate manufacturing), spectroscopic features due to ELC interference (or ``line pulling"), and a voltage offset caused by stray charge accumulation, galvanic potentials, or other stray electric field sources. 
We discuss each of these below.

\begin{figure}[h!]
	\centering
		\scalebox{.4}{\includegraphics*{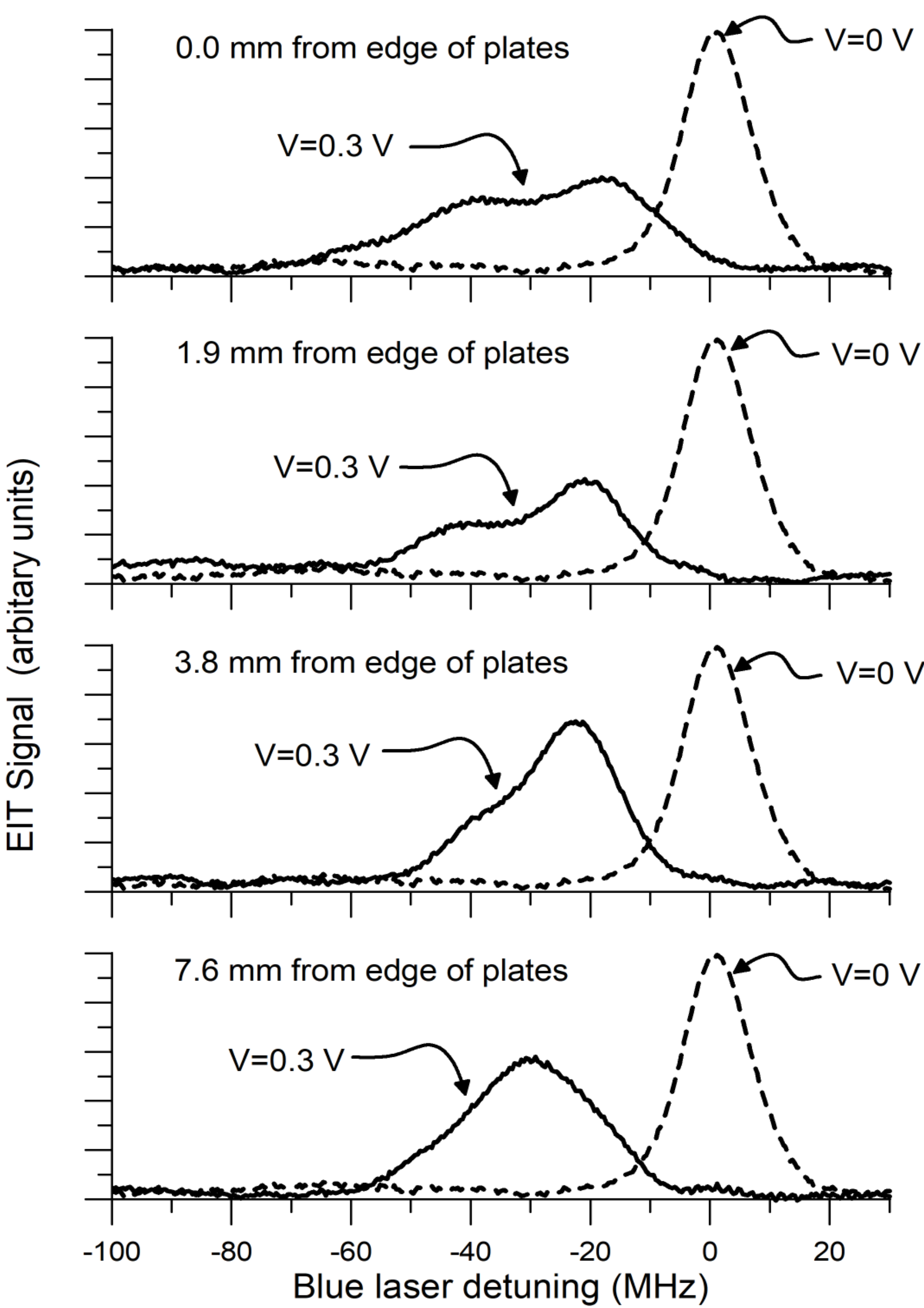}}
	\\
	\caption{EIT signal as the optical beams approach the edge of the plate. The plate has a width of 18~mm, the edge of the plate is define at $x$=0~mm and the center of the plate is at 9~mm from the edge.}
	\label{edge}
\end{figure}

The fringing fields at the edge of the plate can cause a broadening of the EIT line.  Fig.~\ref{edge} shows the EIT signal for an applied voltage of 0.3~V as the optical beams are moved from the center of the plates ($x$=7.6~mm) to the edge of plates ($x$=0~mm). Recall the optical beams are propagating along the the long dimension of the plates. In this figure, we see the EIT signal becomes broader as the beams approach the edge of the plate and we see that the EIT line shape begins to change.  Such effects are due to E-field inhomogeneity in the region near the edge where the fringing fields are present (in this case the beams are approaching two sharp corners of the plates). The angle of the cell relative to the incident angle of the beam will play a large role in the broadening as well.  With that said, such effects can be accounted for in a calibration.  In such calibrations, one needs to be careful when double peaks appear.
These double peaks can be the result of imperfections in the plate manufacturing. 

\begin{figure}[h!]
	\centering
\scalebox{.35}{\includegraphics*{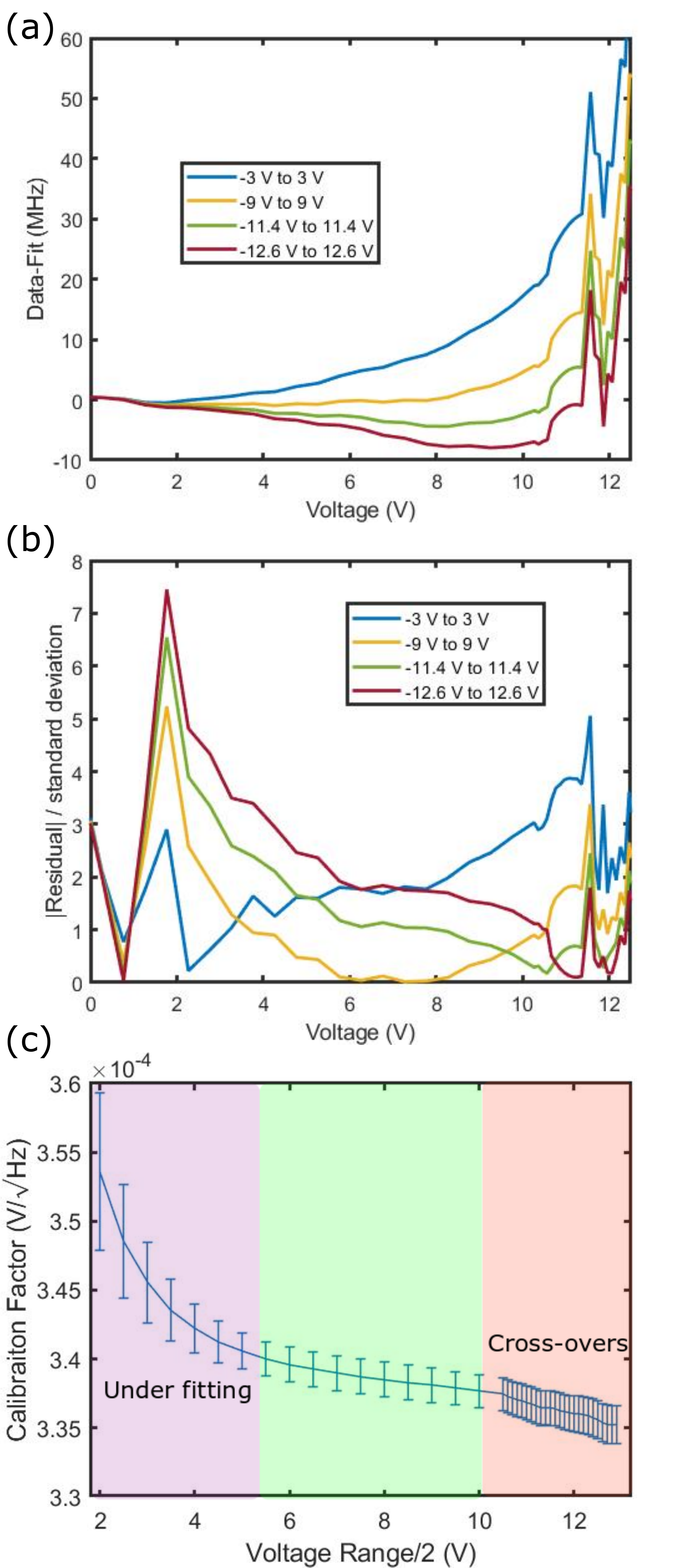}}
	\\
	\caption{(a) Difference (residuals) between the experimental data and the fit in Fig.~\ref{fig:data_shift}. (b) Studentized residuals from (a). Each trace is for a fit incorporating data from the negative voltage out to the positive voltage, as labeled. (c) The calibration factor obtained from fits using different voltage ranges plotted against half the voltage range. This data is for the 28S Rydberg state.}
	\label{fig:subtleties}
\end{figure}

In the case of the ELCs, we analyze their effect on the fit of eq.~(\ref{fit1}). Ideally, the quadratic Stark effect should fit the data well, but the ELCs introduce higher order terms. This can introduce errors in the fit of the data used to find the calibration constant $C_{cal}$. Therefore, we analyzed how the calibration factors and fits change if we included different voltage ranges (max voltage-min voltage, centered at 0 V) during the fitting process. Fig.~\ref{fig:subtleties} shows the difference between the data and the fit [i.e., eq.~(\ref{fit1})] for different voltage ranges given by the separate traces. Fig.~\ref{fig:subtleties}(a) shows the difference (i.e., residuals), Fig.~\ref{fig:subtleties}(b) shows the studentized residuals ($|{\rm residual}|$/standard deviation), and Fig.~\ref{fig:subtleties}(c) shows the calibration factors obtained from the fits for different voltage ranges. Here, we plot the percent difference between the data and the fit for cases where we include different voltage ranges for the different fit traces for the $28S_{1/2}$ state. Note that the large deviations for $<2$~V in Fig.~\ref{fig:subtleties}(b) are a result of the smaller error for the smaller voltages applied. This is due in part to the increase in width of the EIT peak as a voltage is increased. For smaller voltages, the standard deviation was less than 1~mV for the peaks with smaller widths, leading to increased fluctuations.
It can be seen that the fit with the least deviation is for the case where we use voltages from -9 V to 9 V, corresponding to a range of 18 V. In any other case, the fit breaks from the data very quickly. The fits shown in Fig.~\ref{fig:data_shift} and the calibration factor $C_{cal}$ given above were obtained by optimizing the voltage range in this manner.

Also shown in Fig.~\ref{fig:subtleties}(b) are the effects of the ELCs which result in a higher order dependence and do not simply follow the quadratic Stark effect. This is further demonstrated in Fig.~\ref{fig:subtleties}(c), where we see how the calibration factor changes for the different voltage ranges used in the fit. As expected, we see under fitting if not enough data is used (magenta region), a nearly flat line with-in error for the region of good fit (green region), and then a skew of the slope as the voltage range is increased into the region of the ELCs (red region). To overcome the effects of the ELCs at higher voltages, we can use a more precise model to account for their effects, shown by Fig.~\ref{fig:data_shift}. The other option is to tune to different Rydberg states to tune the sensitivity or adjust the plate separation. The latter requires the use of several different cells manufactured for specific operating conditions. By tuning the plates, we change the necessary voltage to produce a given electric field.

\begin{figure}[h!]
	\centering
\scalebox{.5}{\includegraphics*{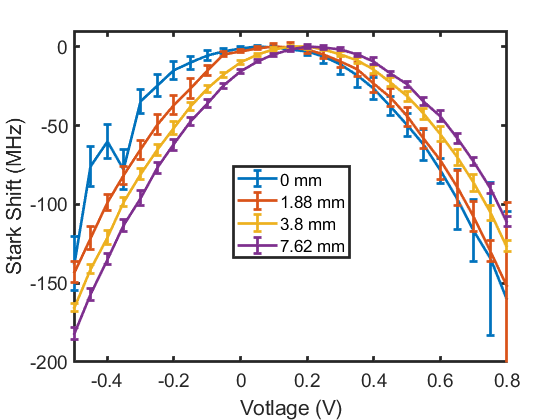}}
	\\
	\caption{Stark shift measurements for different horizontal beam positions parallel to the plates. The edge of the cell is at 0 mm and 7.62 mm is near the center of the cell.}
	\label{fig:subtleties2}
\end{figure}

In the case of the voltage offset, we found that a potential cause is the combination of fringing fields and the charge density present on the plates. As our parallel plates are not infinite, there are contributions which arise from edge effects. These effects have a particularly strong response if the cell is translated horizontally, shown by Fig.~\ref{fig:subtleties2}. In this case, we mean perpendicular to the optical beams and parallel to the plates. As the cell position is moved so that the optical beams are at the edge of the plates (0 mm trace in Fig.~\ref{fig:subtleties2}), we observe that the voltage offset decreases. 


While the positioning of the cell changes the voltage offset, it does not change the shape of the curve. With the exception of the beam position at the very edge of the cell, the calibration factor changed by less than 2~\% for the other measurements. So long as the the calibration curve is acquired and the voltage offset is accounted for, this method has potential as a standard of measurement. However, at the edge of the cell, there is an increased uncertainty in measurement.

Finally, non-uniform plate separation can be an issue for these Rydberg atom sensors. One remedy is to use a microfabricated cell \cite{john2} to insure plate uniformity, where microfabrication of a vapor cell allows for better control of the plate or electrode separation as described in Ref. \cite{john1}. The use of similar vapor cells with internal electrodes are currently being investigated for voltage measurements and other applications \cite{das2}. However, in this type of cell, care needs to be taken to ensure that any coating placed on the electrodes does not cause shielding of the applied voltage and in turn the E-field seen by the atoms~\cite{das1}.



\section{ac Voltage Measurements}

\begin{figure}
	\centering
		\scalebox{.3}{\includegraphics*{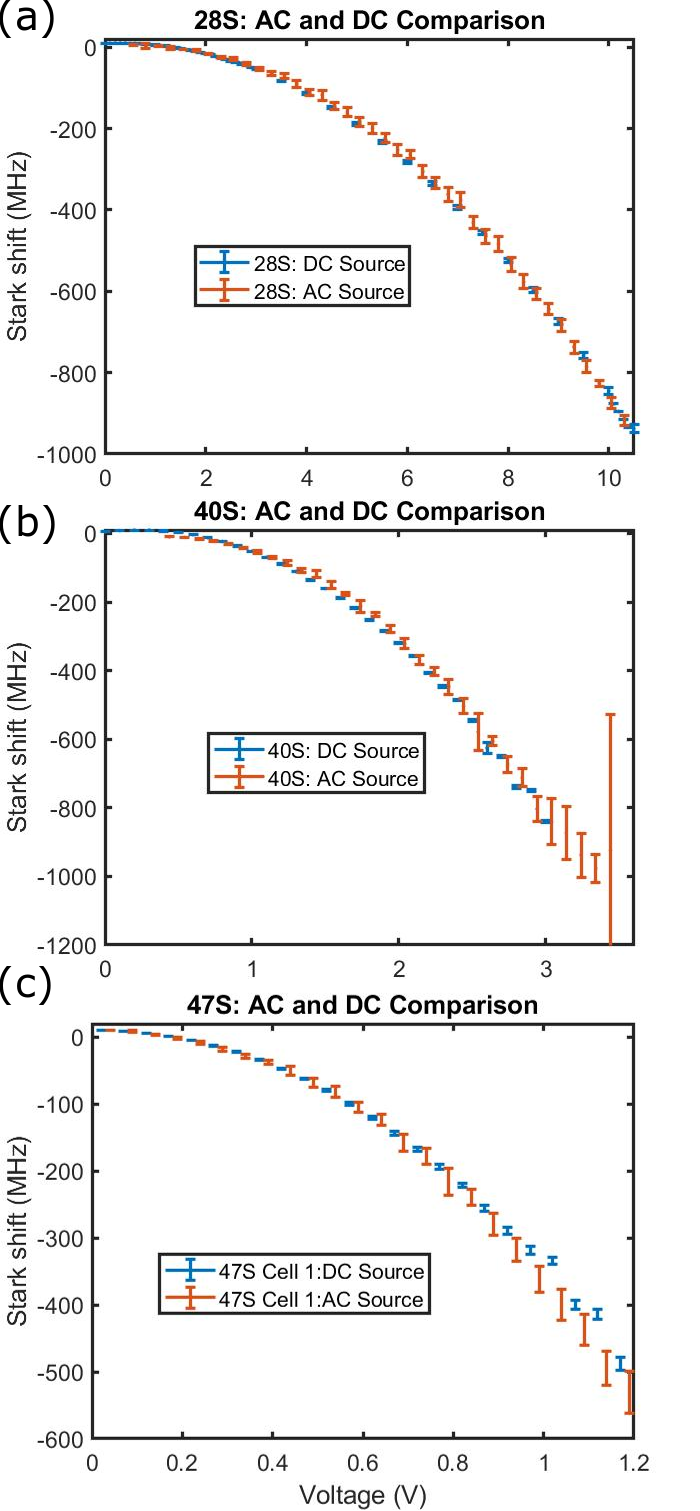}}\\
	\caption{Figure shows the Stark shift plotted against the peak voltage of an 60~Hz ac source for the three states 28S (a), 40S (b), and 47S (c). Also included for comparison are the dc experimental data from Figs.~\ref{fig:data_shift}(a)-(c)}
	\label{fig:AC_data_meth1}
\end{figure}

In addition to dc measurements, this same system can be used to determine 60~Hz ac field strengths. However, 60~Hz ac voltage measurement require different read-out methods. The ac Stark effect would result in a linear shift to the EIT peak, but we instead rely on the slowly changing dc Stark effect in the adiabatic limit where the frequency is slow enough such that the atoms can respond. This limits the measurements to the range of frequencies from dc to 1 MHz; we work with 60 Hz. In this application, we observed that the EIT peak shifts from the 0-voltage location to the peak-voltage location, as expected. Since, the calculated polarizability for the dc and ac fields are nearly identical, this peak voltage location also corresponds to the equivalent dc voltage. So by tracking the maximum EIT peak shift, we can make a calibration curve that matches the curve for a dc source, shown in Fig.~\ref{fig:AC_data_meth1}.

\begin{figure}
	\centering
	\scalebox{.3}{\includegraphics*{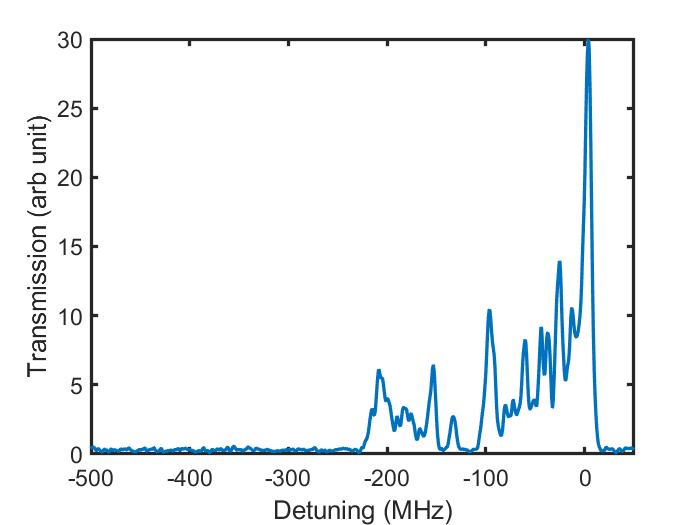}}\\
	\caption{Sample of variance of 20 traces plotted against the coupling laser detuning for a 1.2 V AC source for the 40S state. As the EIT peak shifts around, the variance increases in the regions where it has passed.}
	\label{fig:ac_sample_1}
\end{figure}

For detecting 60~Hz ac volatge, one might consider sweeping a laser fast enough to track the EIT as the 60 Hz source caused it to oscillate from peak-voltage location to zero-voltage location. Unfortunately, such a method requires the laser to scan over a GHz in frequency at a scan frequency much larger than the source bandwidth. Such a scan speed for CW systems results in mode instability and loss in lasing. Here, we demonstrate two alternative methods for determining the peak to peak voltage of the source from the motion of the EIT peak. The first method (Stark shift tracking) relies on similar scans to the ones in the dc voltage measurements. The coupling laser is scanned over a frequency range while 20 traces of the transmission spectrum are gathered. During this time, the EIT peak is oscillating between the 0-voltage location and the peak-voltage location. We take the 20 traces and find the variance at each detuning, shown in Fig.~\ref{fig:ac_sample_1}. Using this data, we find the maximum shift at the location where the data drops off below a threshold value that lies above the noise. The extracted drop-off locations are plotted in Fig.~\ref{fig:AC_data_meth1} for the three different states also used for the dc measurements. Also plotted are the dc data to show the agreement in the calibration between the dc data and the ac data taken with this method.

Unfortunately, there is more noise in the ac data than in the dc data. This noise is due to the accuracy of determining the drop-off location, which becomes increasingly more difficult for higher ac voltages. At higher voltages, the drop-off almost smooths to the level of the noise. The dc data did have a similar issue for the higher voltages, but because of the averaging of the peak at one location, it was mitigated. With an ac source, this option is not available and for the interest of time and as a demonstration, we only used 20 traces. However, by collecting more traces, we should be able to bring the noise down. Further sources of error or deviation from the calibration could also be present from the imperfections in the plates of the cell. The EIT peak widths broadens with voltage and this could potentially result in a voltage dependent shift in the drop-off location. However, this is not apparent here within the bounds of error.

\begin{figure}
	\centering
	\scalebox{.6}{\includegraphics*{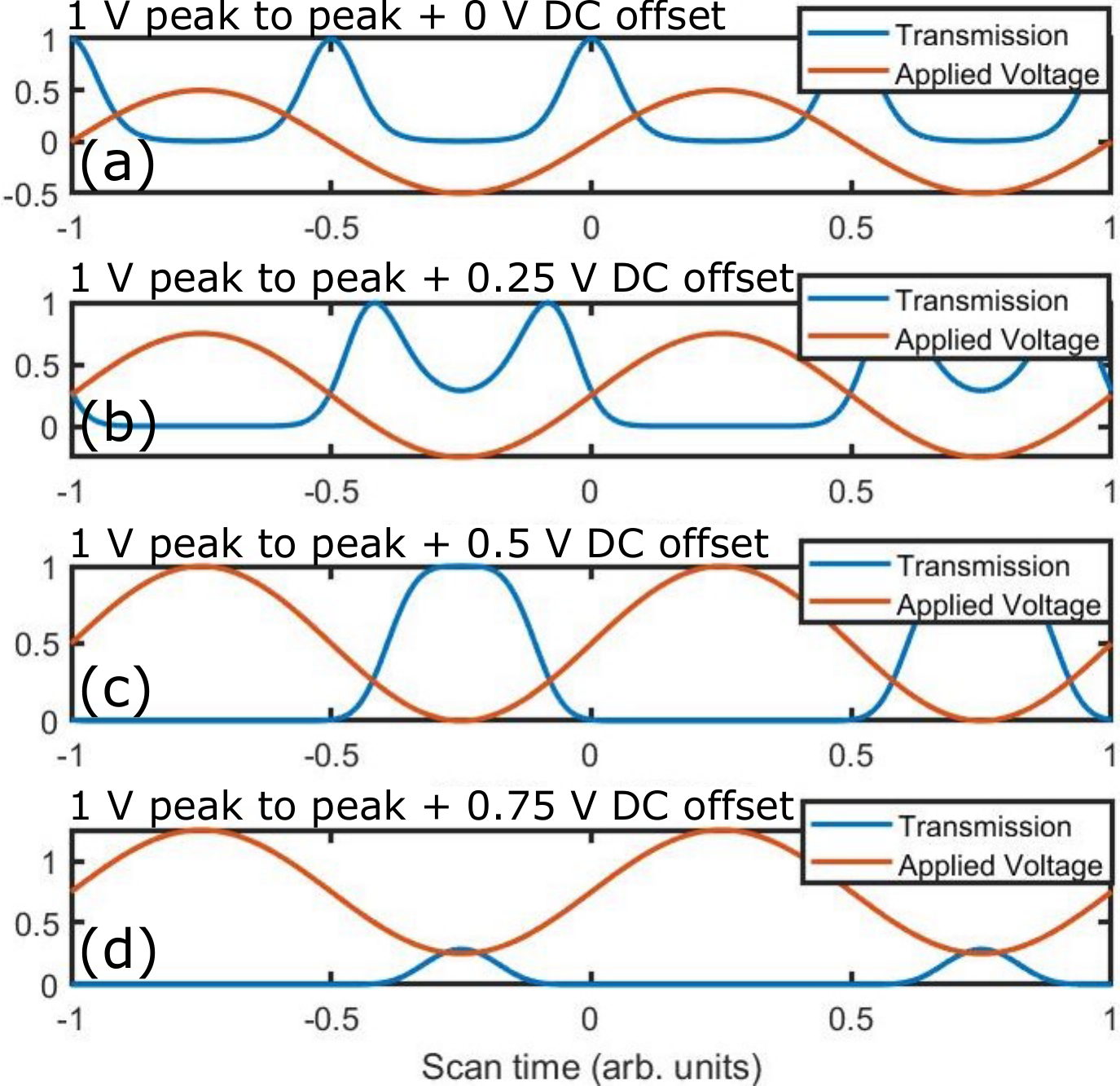}}\\
	\caption{Sample showing expected EIT transmission (blue) in the presence of a 60 Hz ac field (red) plotted against the phase of the applied ac field. The plots correspond to offset voltages of (a) 0 V, (b) 0.25 V, (c) 0.50 V, and (d) 0.75 V, as labeled.}
	\label{fig:AC_sample_2}
\end{figure}

The second method (dc biased ac sensing) relies on the use of a dc calibrated source. As discussed previously, if we have an ac source on the plates, the EIT peak will oscillate between the zero-voltage location and the peak-voltage location. We can observe this shifting if we scan the coupling laser. However, if we lock the coupling laser to the zero-voltage location, we will instead observe transmission only when the ac voltage passes through the zero-voltage location, as shown by the computer generated plot in Fig.~\ref{fig:AC_sample_2} (a). This produces a transmission peak that we can monitor that is dependant on the ac voltage source. Now, if we apply a calibrated dc voltage along with the ac voltage the transmission peak will shift since the ac voltage at which the voltage sum crosses zero is no longer when the ac voltage passes zero, shown in Figs.~\ref{fig:AC_sample_2} (b) and (c). However, if the dc voltage is increased to a value higher than the peak voltage, the transmission will begin to vanish, as shown in Fig.~\ref{fig:AC_sample_2} (d). 

\begin{figure*}
\centering
\scalebox{.58}{\includegraphics*{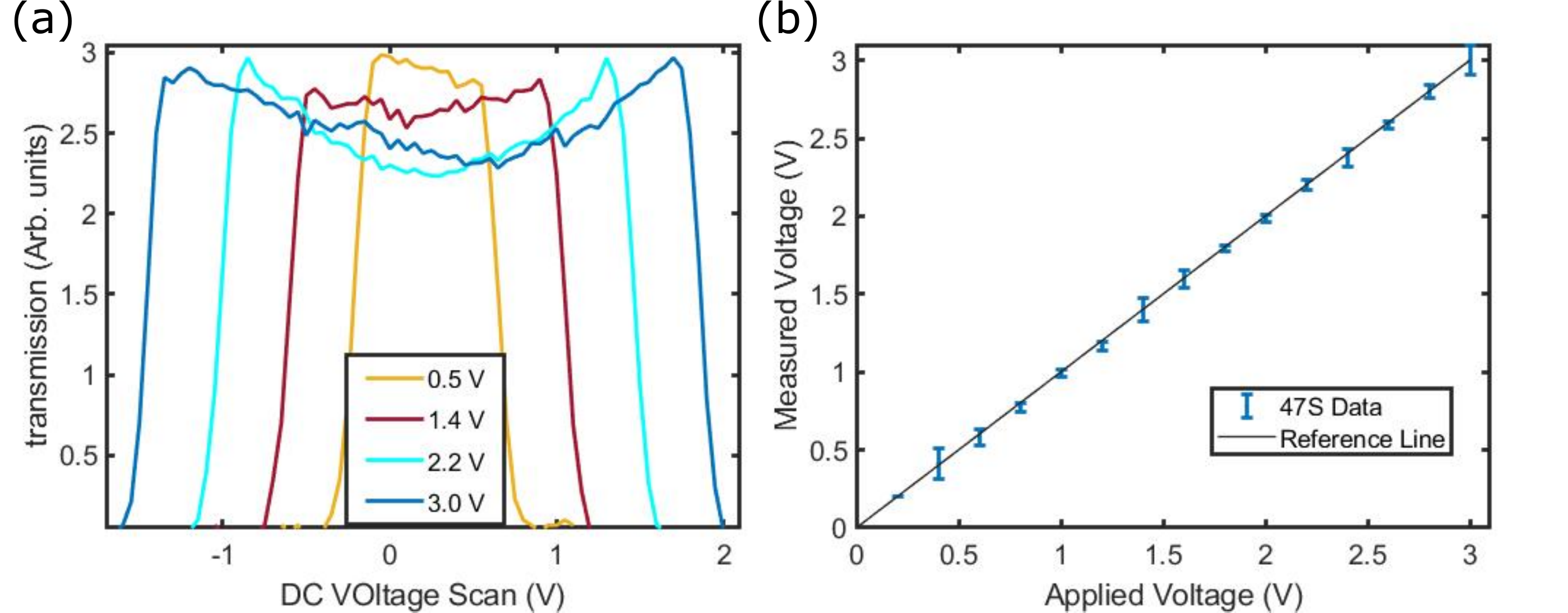}}\\
	\caption{(a) Sample of traces for different ac voltages where the peak transmission is plotted against the calibrated dc voltage. (b) The width of the traces in (a) that correspond to the peak to peak voltage, plotted against the applied peak to peak voltage (V). This data is for the 47 S state. (black) is the one-to-one line as a guide and the (blue) is the extracted ac voltage. }
	\label{fig:AC_meth2}
\end{figure*}

In this method, we apply a calibrated dc voltage as an offset to an ac voltage source that we wish to determine. We track the height of the transmission peaks in Fig.~\ref{fig:AC_sample_2}. As we scan the dc voltage source from -2~V to 2~V, we can see the transmission appear and then vanish, as shown in Fig.~\ref{fig:AC_meth2} (a). The voltage difference in the two drop-off locations in the dc voltage scan defines the peak to peak voltage of the applied ac voltage. Fig.~\ref{fig:AC_meth2} (a) shows the traces of dc voltage scan for several different ac voltages and Fig.~\ref{fig:AC_meth2} (b) shows the extracted peak to peak voltages plotted against the applied voltage. 

DC biased ac sensing has certain advantages when compared to Stark shift tracking. In the case of Stark shift tracking, we must average for a long time to improve the statistics enough to measure larger voltages. However, for dc biased ac sensing, we do not have this same limitation since the measurement occurs at the zero-voltage location with maximum EIT peak height. Unfortunately, the maximum ac voltage we can sense is limited by the range of our calibrated dc source. While this voltage is higher than measurements with the Stark shift tracking method, it is still limited. Furthermore, this method requires a laser lock that is stable to less than 1 MHz for the probe and coupling lasers.

\section{Conclusion and Discussion}
In this manuscript, we have demonstrated a means to an alternative voltage standard based on the inherent and known dipole strengths of Rydberg atoms and have filled the SWAP-C gap between Zener diode and Josephson junction based standards in doing so.
By measuring the dc Stark shift for various voltages at different states, we generate a calibration standard which can be utilized as an alternative to two established voltage measurement technologies. In addition to this, we discuss how the calibration factor $C_{cal}$ can be used in two ways. First, if the plate separation is known, the calibration factor can be used to determine the polarizaibility for the Rydberg states. Secondly, we discuss how calibration factors can be used to find the plate separation to within 10s of microns. We also demonstrate two methods to measure 60~Hz ac sources with little to no modification to the apparatus. 

In the case of dc sources, we were able to find Stark shifts ($\Delta$) within an average error under $\pm 5$~MHz over 6 sets of data. Even though there was broadening of the EIT peak and contributions from inhomogeneities at larger voltages, the uncertainty in the voltage did not change since the
sensitivity increases with larger voltages. The fractional uncertainty in the voltage is given by
\begin{equation}
\left(\frac{\delta V}{V}\right)^2 =\left(\frac{\delta d}{d}\right)^2 + \left(\frac{\delta\alpha}{2\cdot\alpha}\right)^2 +\left( \frac{\delta\Delta}{2\cdot\Delta}\right)^2.
\end{equation}
The error in the Stark shift measurement translated to a fractional uncertainty in the voltage is on the order of ~.01 V/V on average over the voltage range of the 47S Rydberg state. The fractional uncertainty is constant due to the broadening of the EIT peak at higher applied voltages, as discussed previously. The 20 sets averaged to find a given Stark shift have scan times of 25 ms. This corresponds to a sensitivity in the voltage measurement on the order of 0.007 V/V\textrm{$~Hz^{-1/2}$} using the 47S state. While this level of sensitivity seems limited, different Rydberg states can be used for increased sensitivity. By simply choosing to work at the 28S state, the uncertainty in the measurement decreases to 0.003 V/V. This is due to the stronger EIT peak at the 28S Rydberg state. The sensitivities of these voltage measurements are ultimately limited by several aspects. One is that the polarizability can only be calculated to an uncertainty of 1\%. However, certain experimental techniques can determine the polarizabilities of atoms to a much better accuracy~\cite{bai_2020}. Another key factor is in the time domain. The laser sweep must be able to span several GHz for these measurements and thus sets a limit on how fast data can be collected. 
Finally, the ultimate determination of how well the Stark shift can be measured depends on the linewidth and amplitude of the EIT line. This is currently 15~MHz in our system. Similar Rydberg techniques have demonstrated linewidths down to 2 MHz~\cite{kumar1}. This is nearly an order of magnitude improvement that can be achieved. Furthermore, new three-photon configurations may offer linewidths under 100~kHz which could offer over an order of magnitude improvement~\cite{shaffer_three}. In addition to this, for increased voltages, we begin to see effects from energy level crossings and plate inhomogeneities. The prior can be addressed by adjustments to the quadratic fit and the latter by the manufacturing of a more uniform plate geometry, as our current plates have a dimple (and discontinuities) in the very center where the leads are mounted. These improvements are complementary and would increase the range of voltages accessible and improve our sensitivity to better than 10\text{$^{-5}V/V~Hz^{-1/2}$}.


We analyzed various sources of uncertainty that arose from non-uniform electric fields, ELCs, and plate imperfections. The combination of these effects initially point to the infeasibility of the Rydberg atom as a voltage standard, but through careful consideration and calibration, these problems can be accounted for. We discussed the broadening of the EIT lines due to non-uniform fields, EIT line perturbations from fringing fields at the edge of the plates, and the effects of the ELCs on the EIT spectra. These considerations are key to developing a portable Rydberg standard and will affect the sensitivity, range, and reliability for future atom based voltage sensors. In conclusion, this study on the dc and 60 Hz ac fields will provide insight for future advancements for the realization of a self-calibrating, deployable voltage standard.  Future work will include developing vapor cells with more precise plate separation, including microfabricated cells.

We emphasize that a future device could be made where instead of being calibrated, the polarizability and plate spacing could be accurately measured and then the device would be based on fundamental standards. Other questions to be answered in future work is how the resolution, stability and fractional uncertainties compare to Zener diodes systems.







\section{Acknowledgement}
The authors thank Dr. Yuan-Yu Jau with Sandia National Labs, Albuquerque, NM 87123, USA for his useful, insightful technology discussions, and for suppling one of the vapor cells used in our experiments.

\section{Data Availability Statement}
\vspace{-5mm}\noindent Data is available upon request.

\end{document}